\documentclass[fleqn,usenatbib]{mnras}

\usepackage[T1]{fontenc}

\DeclareRobustCommand{\VAN}[3]{#2}
\let\VANthebibliography\thebibliography
\def\thebibliography{\DeclareRobustCommand{\VAN}[3]{##3}\VANthebibliography}

\usepackage{graphicx}	% Including figure files
\usepackage{amsmath}	% Advanced maths commands
\usepackage{pdflscape}
\usepackage{threeparttable}

\title[Three Cases of Optical Periodic Modulation in AGN]{Three Cases of
Optical Periodic Modulation in Active Galactic Nuclei}

\author[Li et al.]{
Jie Li,$^{1}$
Zhongxiang Wang,$^{1,2}$\thanks{E-mail: wangzx20@ynu.edu.cn}
and Dong Zheng$^{1}$
\\
$^{1}$Department of Astronomy, School of Physics and Astronomy, Yunnan University, Kunming 650091, China\\
$^{2}$Shanghai Astronomical Observatory, Chinese Academy of Sciences, 80 Nandan Road, Shanghai 200030, China
}

\date{Accepted XXX. Received YYY; in original form ZZZ}

\pubyear{2022}

\begin{document}
\label{firstpage}
\pagerange{\pageref{firstpage}--\pageref{lastpage}}
\maketitle

\begin{abstract}
	We report on the case of optical periodic modulation discovered in two Active
	Galactic Nuclei (AGN) and one candidate AGN.
Analyzing the archival optical data obtained from large transient surveys,
	namely the Catalina Real-Transient Survey (CRTS) and the Zwicky 
	Transient Facility (ZTF), we find periodicities 
	of 2169.7, 2103.1, and 1462.6\,day in sources J0122+1032,
	J1007+1248 (or PG~1004+1248), and J2131$-$1127, respectively. 
	The optical 
	spectra of the first two indicate that the first is likely a blazar and 
	the second a type 1 Seyfert galaxy, and while no spectroscopic information
	is available for the third one, its overall properties
	suggest that it is likely an AGN. In addition, mid-infrared (MIR) 
	light curve data of the three sources, taken by the Wide-field Infrared
	Survey Explorer (WISE), are also analyzed. The light curves show 
	significant variations, but not appearing related 
	to the optical periodicities.
	Based on the widely-discussed super-massive black
	hole binary (SMBHB)
	scenario, we discuss the origin of the optical modulation.
	Two possible interesting features, an additional
	162-day short optical periodicity in J2131$-$1127 and the consistency of
	the X-ray flux variations of J1007+1248 with its optical periodicity,
	are also discussed within the SMBHB scenario.
\end{abstract}

\begin{keywords}
	Quasars:supermassive black holes 
\end{keywords}

%%%%%%%%%%%%%%%%%%%%%%%%%%%%%%%%%%%%%%%%%%%%%%%%%%

%%%%%%%%%%%%%%%%% BODY OF PAPER %%%%%%%%%%%%%%%%%%

\section{Introduction}
Enabled by the availability of huge amounts of data collected from large 
survey programs 
conducted in recent years, searches and studies of long-term variabilities of
different types of sources become possible. The Active Galactic nuclei (AGN)
are naturally the focus of these studies, as their variability timescales 
can be long, driven by activities of the super-mass black holes (SMBHs)
at their center. Related phenomena include the so-called quasi-periodic 
oscillations (QPOs), a fraction of which 
have year-long periods (e.g., see \citealt{zw21} and references therein)
and have been discussed to reflect the binary SMBH nature 
for the reported AGN systems (see, e.g., \citealt{val+08,ack+15,sss17}).
Another similar and probably more interesting phenomenon is the optical 
periodicities, revealed in large part by recent large optical 
transient surveys. 
In the Catalina Real-time Transient Survey (CRTS; \citealt{crts09}),
111 quasars were reported 
to show candidate periodicity signals in 9-yr long light curves \citep{gra+15b},
in addition to the quasar PG~1302$-$102 that was found to show 
a $\sim$1884\,day periodicity \citep{gra+15a}.
By systematically analyzing the Medium Deep Survey
data obtained from the Panoramic Survey Telescope and Rapid Response System 
(Pan-STARRS), \citet{liu+15} and \citet{liu+19} respectively reported 
a possible periodic signal of $\sim$542\,day from a luminous radio-loud
quasar (PSO J334.2028+01.4075) and 26 candidates of similar
periodic patterns. Also, by searching through data 
from the Palomar Transient Factory (PTF), 
\citet{cha+16} reported 50 candidates with significant periodicities,
33 of which remained significant
when the CRTS data were added in the analysis.

These long-term periodicities have been widely discussed as signals indicating
the existence of SMBH binaries (SMBHBs) situated at the center of AGN. 
According to
hierarchical models of galaxy formation and evolution, galaxy mergers
are common in their formation history (e.g., \citealt{del+06}), and as a result,
a significant fraction of galaxies would contain SMBHBs at their center
(e.g., \citealt{vhm03}). Such a binary system would go through complicated 
evolutionary processes, ending with coalescence driven by gravitational
radiation (see, e.g., \citealt{col14} for a review). Over the course 
of the evolution,
the dominant time period ($\sim$Gyr) the SMBHB spends at would be
when the two SMBHs have a sub-parsec separation distance
(i.e., hardening phase; e.g., \citealt{yu02,hkm09,col14}). 
Depending on the masses of the binary system,
the corresponding orbital period could be several years.
Therefore,
the periodicity signals detected from the surveys match the theoretical 
expectations, although
how the signals are produced is not totally clear \citep{gra+15a}. Different
numerical simulations for the SMBHB systems have shown that each black hole
in such a binary, surrounded by a circumbinary disc, would have a 
so-called mini-disc
due to the accretion supplied by the mass transfer from the circumbinary disc.
The accretion rate of the black holes would 
be modulated at the orbital period (e.g., \citealt{hms07,mm08,cua+09,far+14}). 
The presence of periodic flux variations would thus be related to the orbital
modulation of the accretion rate \citep{hkm09}.

The Zwicky Transient Facility (ZTF) survey, starting from 2018 March, represents
a more powerful optical transient survey due to its high cadence in
covering a given
sky and fast data releases \citep{bel+19}. We explored its released data and
tested to search for periodic signals from AGN. In essence, we queried
the ZTF public database for AGN light curves using the Automatic Learning 
for the Rapid Classification of Events (ALeRCE; \citealt{for+21,san+21}),
for which we required $\geq$100 data points in any light curve to ensure 
a clear view of any variation patterns.
We went over the output light curves by eye and selected those showing
long-term trends in the $\sim$4\,yr ZTF-survey time period.
We then checked the CRTS light curve data for possible 
connectable patterns to the ZTF long-term trends. In this way, we were able
to find approximately 30 AGN sources that showed potential long-term
periodic variations in an approximately 6000-day time period, 
set by the CRTS and ZTF data. Among them, there are three good cases.
Possibly related variations for the three cases were also seen
in the mid-infrared (MIR) data obtained with the Wide-field Infrared Survey
Explorer (WISE; \citealt{wri+10}). Here we report on our findings of 
the three
cases. One is likely a blazar and the other two are a known AGN and a candidate 
AGN, previously not known with the periodicity 
phenomenon. In Section~\ref{sec:sd}, we describe
the known properties of the three sources and the available data for them.
In Section~\ref{sec:res}, the periodicity analysis 
method is described, and the analyzing results are presented.
We discuss and summarize the results in Section~\ref{sec:dis}.

\begin{figure}
\includegraphics[width=\columnwidth]{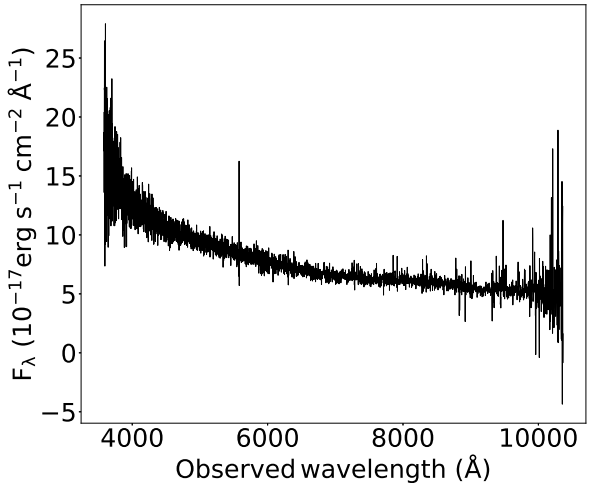}
\caption{Optical spectrum of J0122+1032 from the SDSS database. The absence of
	any line features strongly suggests that the source is a blazar.}
    \label{fig:01spec}
\end{figure}

\section{Sources and data}
\label{sec:sd}

\subsection{J0122+1032}

Little information is available for J0122+1032 (hereafter J0122), but there is 
an optical spectrum, taken on 2018 Nov. 29 (MJD 58451),
in the Sloan Digital Sky Survey (SDSS) database (Figure~\ref{fig:01spec}).
The spectrum shows no obvious features, which indicates the source is likely a
BL Lac type blazar.

The source's coordinates in ZTF, as well as the object ID (oid),
are provided in Table~\ref{tab:data}. We used ZTF's light curve data 
\citep{mas+19}.
In order to ensure the cleanness and goodness of the data, we set
{\tt catflags=0} and {\tt chi$<$4} when querying (the same requirements
were set for the ZTF data of the other two sources), where
the first parameter flags the photometric/image quality and the second
is the root-mean-square metric from point-spread-function fit
(see {\tt The ZTF Science Data System Explanatory Supplement}\footnote{https://www.ztf.caltech.edu/ztf-public-releases.html}).
The light curve data at the ZTF bands $zg$ and $zr$ \citep{bel+19} are
in a magnitude range of 19--20, and the time range of the data is given in 
Table~\ref{tab:data}.

We also obtained the source's CRTS $V$ band data, for which the object
IDs in CRTS and time 
range are given in Table~\ref{tab:data}. The magnitude range is 
between 19--20.

The source was detected with WISE at the W1 (3.4\,$\mu$m) and W2 (4.6\,$\mu$m)
bands in the NEOWISE Post-Cryo survey phase. We downloaded 
the magnitude measurements in the two bands from the NEOWISE-R Single-exposure
Source database.

We searched for X-ray observations that might cover the source, but 
did not find any in the database of the major X-ray telescopes. However,
we note a $\gamma$-ray source 4FGL~J0122.4+1034, detected with the 
Large Area Telescope (LAT) onboard {\it the Fermi Gamma-ray Space Telescope
(Fermi)}, has a position $\simeq$2.2$\arcmin$ away from that
of J0122, while its 95\%-confidence error circle is $\simeq$2.5$\arcmin$. 
Thus this $\gamma$-ray source could be associated with J0122.
It is faint at $\gamma$-rays, having a test statistic
(TS) value of 32.2 (implies a $\sim5\sigma$ detection 
significance). Its emission is described with a power law (the photon index
$\Gamma=1.91\pm0.21$). This source is not identified with a type
or associated with any known source in the {\it Fermi} LAT catalog \citep{dr3}.
We tested to construct a 90-day binned light curve, 
from which flare-like events might be found and the blazar nature would
be identified (e.g., \citealt{wan+20}) since the dominant sources detected 
with {\it Fermi} LAT are blazars \citep{dr3}.
However no obvious flaring-type events were seen.
Whether the $\gamma$-ray 
source is associated with J0122 remains to be investigated.
\begin{figure*}
	\includegraphics[width=0.48\textwidth]{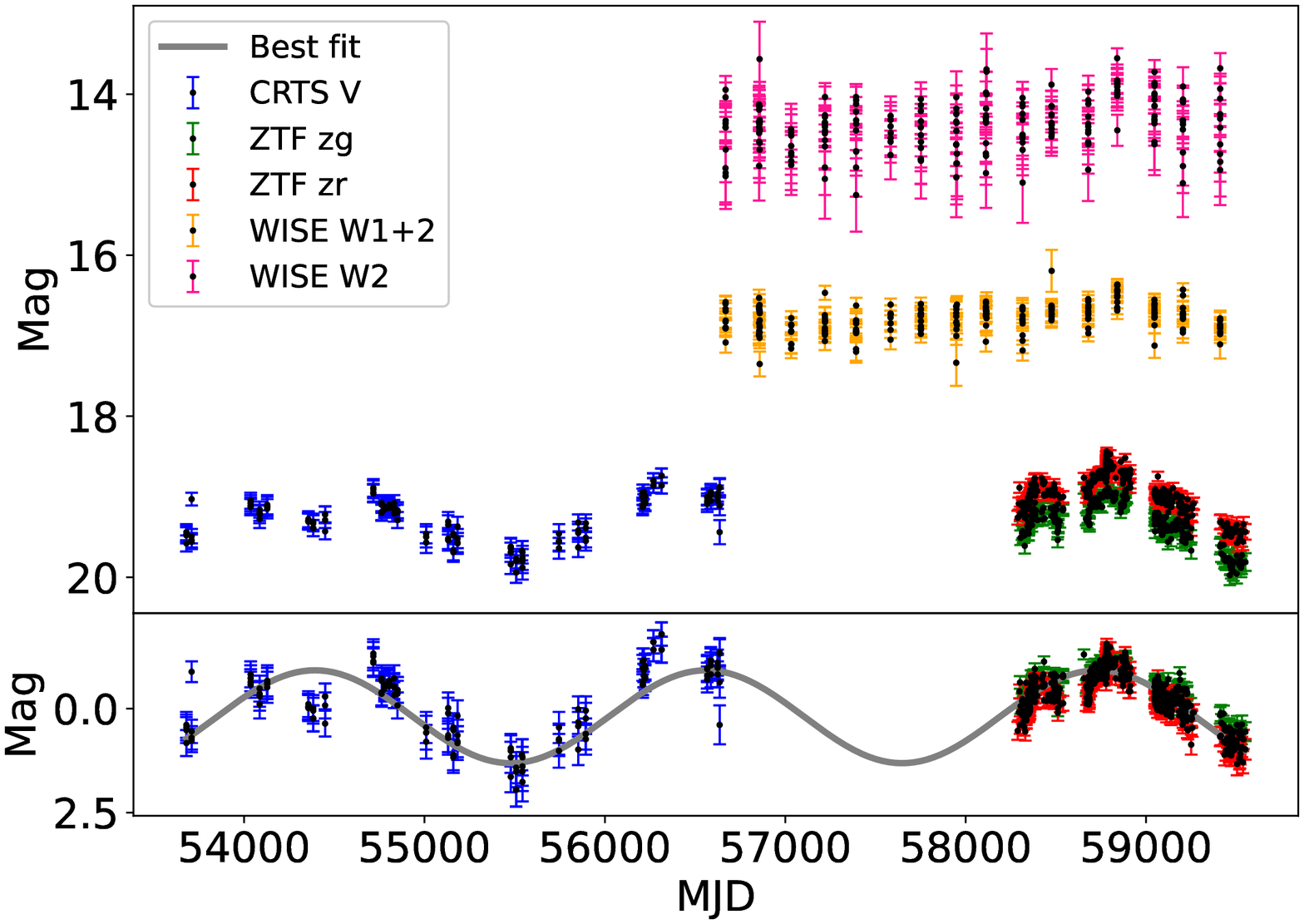}
	\includegraphics[width=0.48\textwidth]{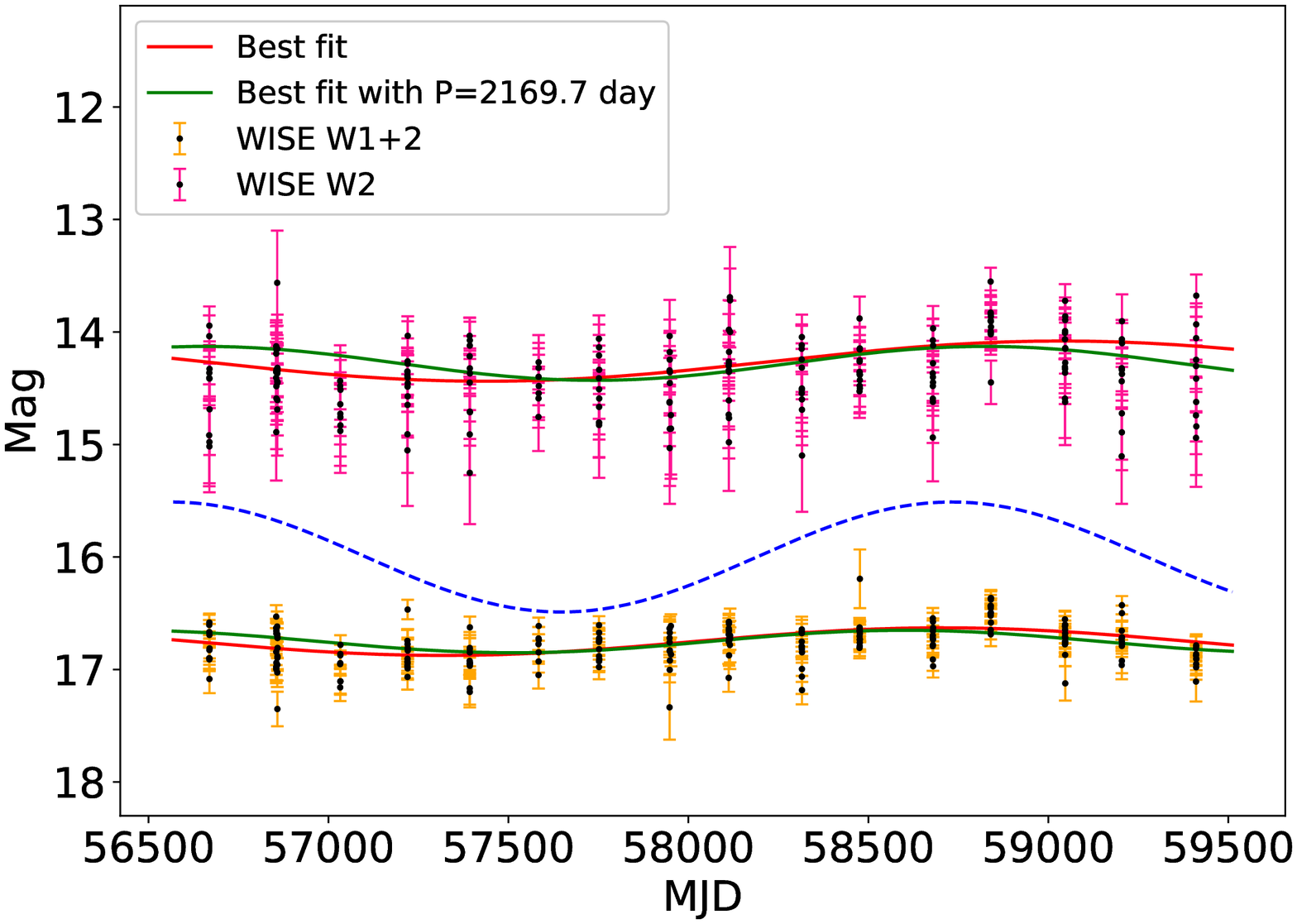}
	\caption{{\it Left panels:} Optical light curve of J0122+1032.
	The WISE W1 (down-shifted by 2 mag for clarity) and W2 band
	measurements are also shown in the {\it top} panel.
	The normalized optical light curve is shown in the {\it bottom}
	panel, which can be well described  with a sinusoidal fit (gray curve).
	{\it Right panel:} Sinusoidal fits to the WISE W1 and W2 band light 
	curves (where the W1 band is down-shifted by 2 mag for clarity), of
	which the green curves are the ones with the period fixed at 
	the optical value. The blue dashed curve is the model fit to the optical
	light curve, shown for comparison.}
    \label{fig:j01}
\end{figure*}

\begin{figure*}
	\includegraphics[width=0.48\textwidth]{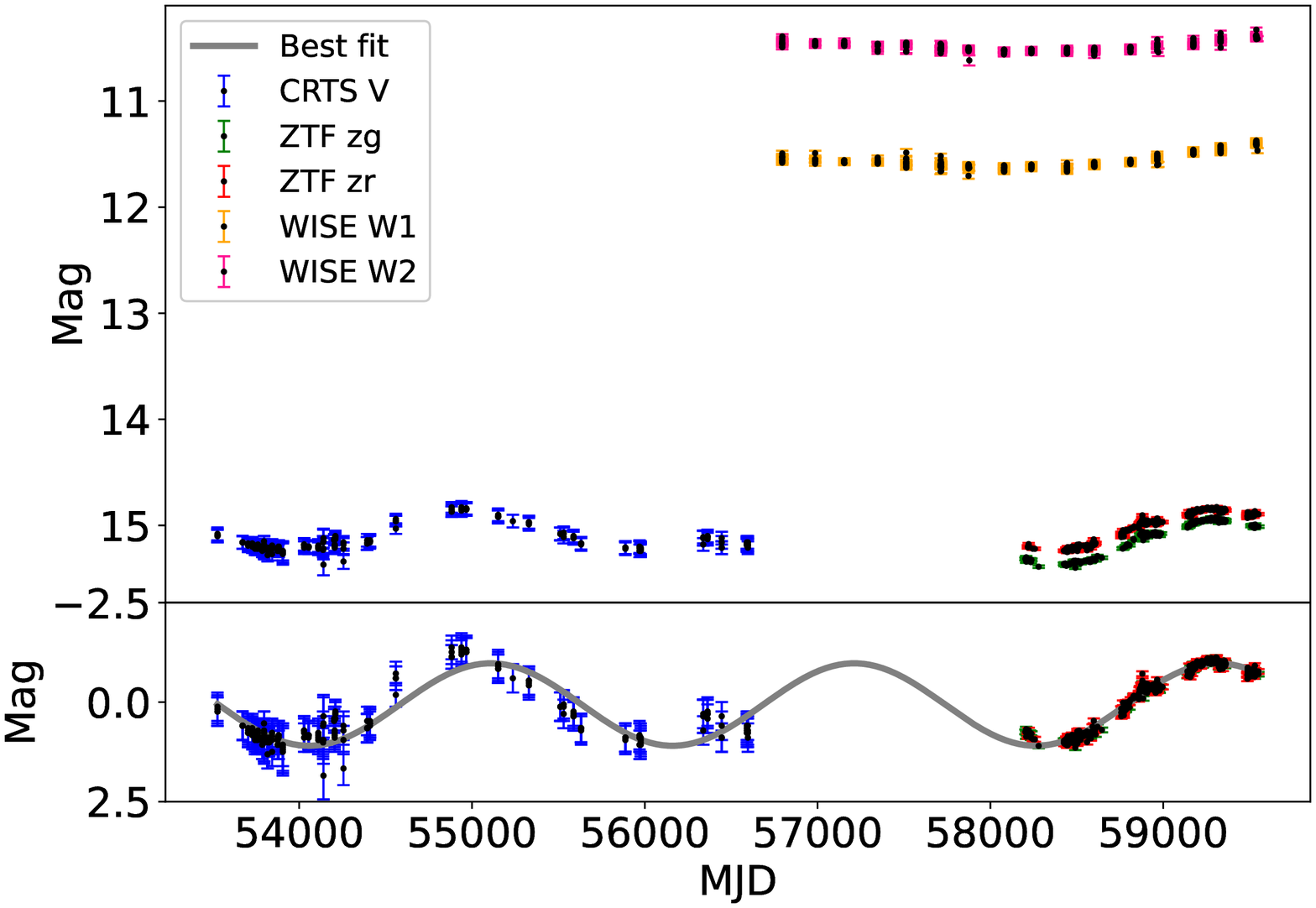}
	\includegraphics[width=0.48\textwidth]{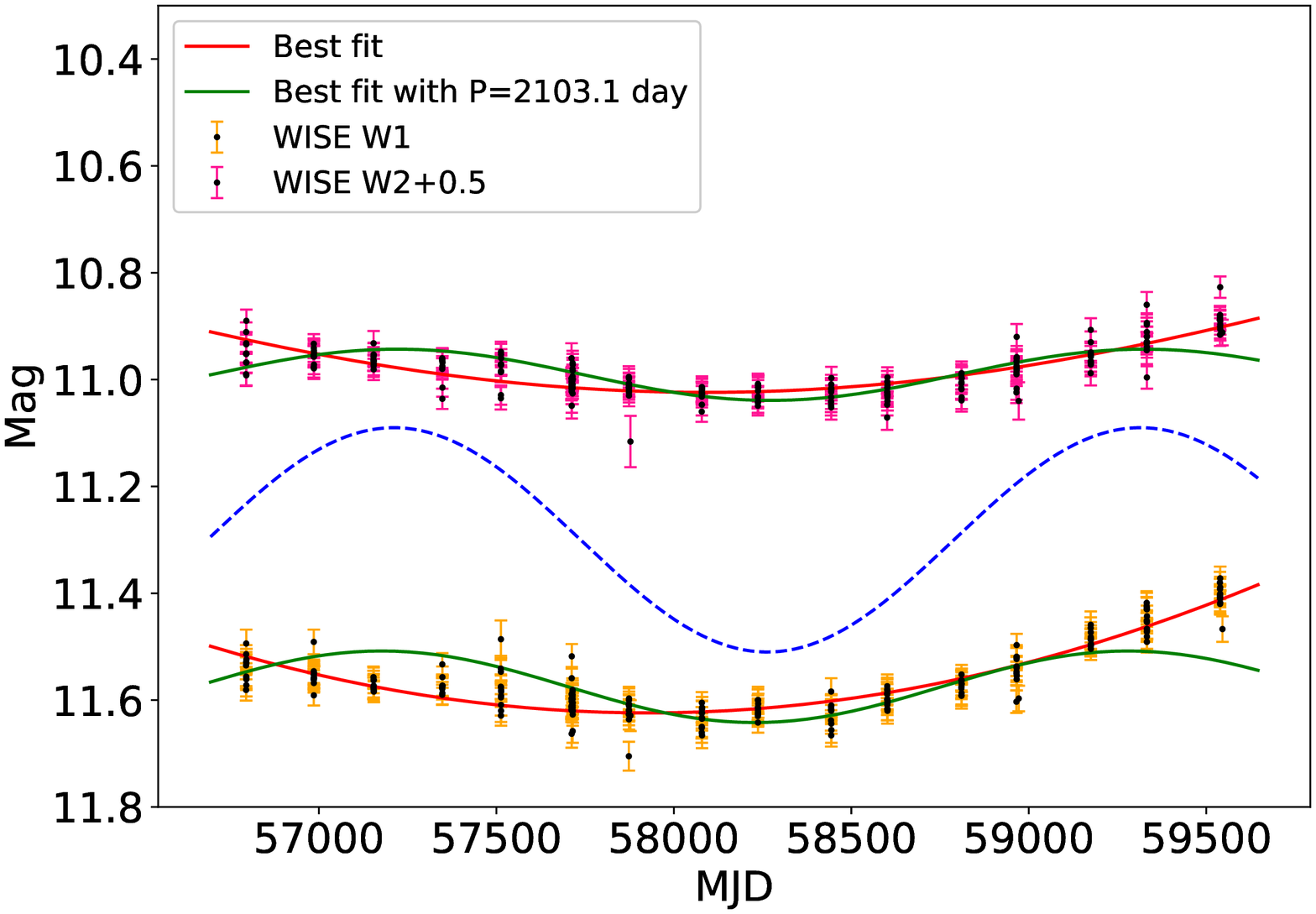}
	\caption{Same as Figure~\ref{fig:j01} for J1007+1248. The W2 band
	light curve in the {\it right} panel is down-shifted by 0.5\,mag for
	clarity.}
    \label{fig:j10}
\end{figure*}
\begin{figure*}
	\includegraphics[width=0.48\textwidth]{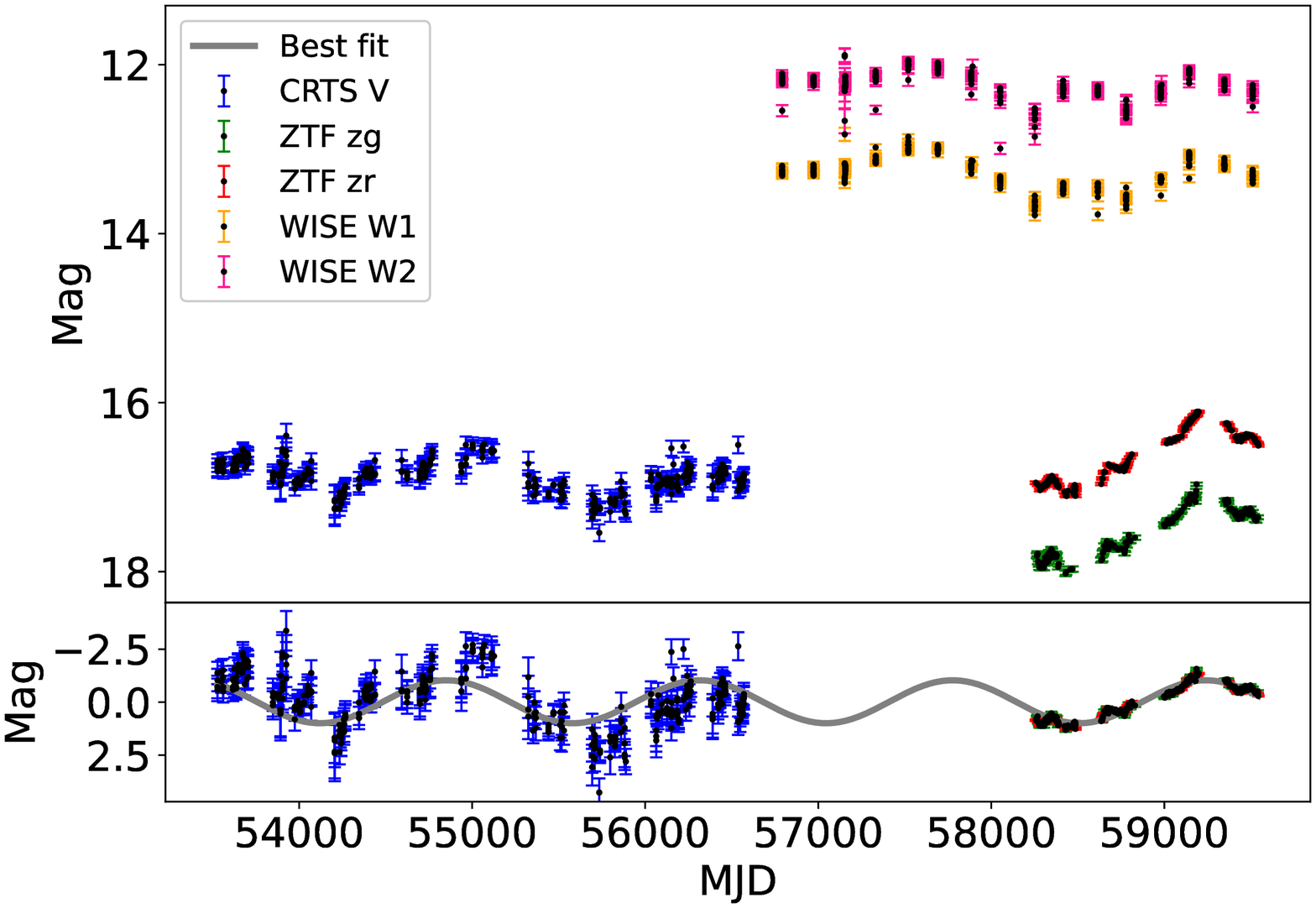}
	\includegraphics[width=0.48\textwidth]{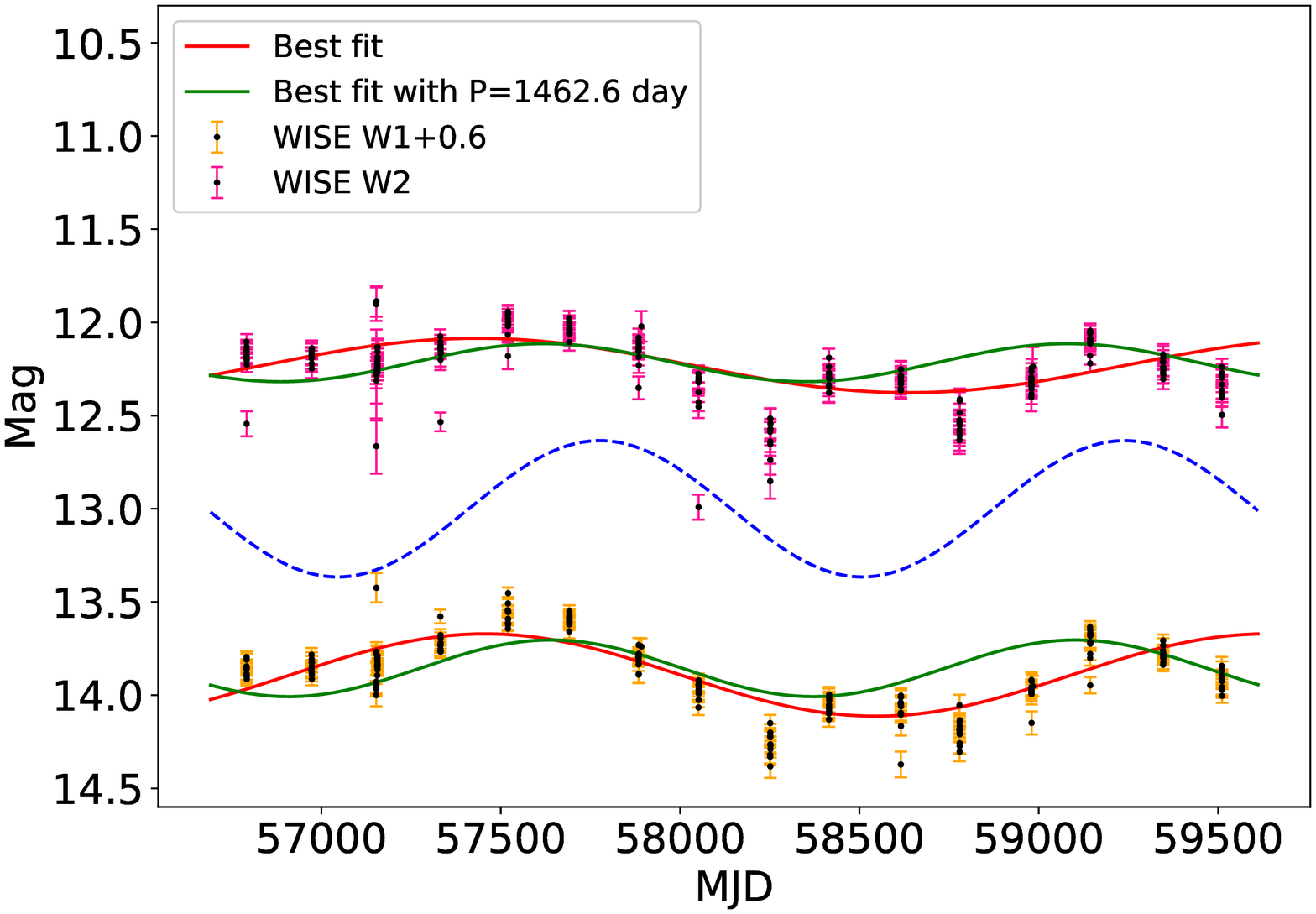}
	\caption{Same as Figure~\ref{fig:j01} for J2131$-$1127. The W1 band
	light curve in the {\it right} panel is down-shifted by 0.6\,mag
	for clarity.
        }
    \label{fig:j21}
\end{figure*}

\begin{table*}
	\centering
	\caption{CRTS, ZTF, and WISE data for the three AGN}
	\label{tab:info}
	\begin{tabular}{lccccccc} 
		\hline
		Source & Coordinate &CRTS ID &ZTF oid & CRTS & ZTF & WISE \\
		       & (RA, Dec) &  &  & (MJD) & (MJD) & (MJD)\\ 
		\hline
		J0122+1032 &(01:22:23.63, +10:32:13.30) &MLS\_J012223.5+103213 &502108400001209  & 53680--56636 &58288--59550 & 56668--59411\\
		           &                         &                      &502208400002496  &            &            &           \\
		J1007+1248 & (10:07:26.11, +12:48:56.20) &CSS\_J100726.1+124856 &1566102100003309 & 53527--56596 &58202--59550 &56794--59545\\
		           &                         &MLS\_J100726.1+124856 &520110100002864  &            &            &           \\
		           &                         &                      &520210100007454  &            &            &           \\
		           &                         &                      &1566202100004632 &            &            &           \\
		J2131$-$1127 & (21:31:06.96, $-$11:27:25.20) &CSS\_J213107.1$-$112724 &390107400004429  &53528--56570 &58257--59543 &56790--59511\\
		           &                         &MLS\_J213107.1$-$112725 &390207400009360  &            &            &           \\
				   &                         &SSS\_J213107.0$-$112724 &                 &            &            &           \\
		\hline
	\end{tabular}
	\label{tab:data}
\end{table*}
\subsection{J1007+1248}

The source J1007+1248 (hereafter J1007), known as PG~1004+130 
(or 4C~13.41), is a radio loud quasar \citep{wbl99} 
and has been relatively well studied. There is an optical spectrum of it,
taken on 2004 Feb. 20, in the SDSS database (SDSS~J100726.10+124856.2)
and it has been identified as a type 1 Seyfert galaxy 
(e.g., \citealt{tob+14}). It has redshift $z\simeq 0.24$, and 
based on the full width at half maximum of the broad H$\beta$ line and 
the optical continuum luminosity, was estimated to
contain a SMBH with mass $M_t = 1.87\times 10^9\,M_{\odot}$ 
\citep{vp06}. \citet{sco+15} conducted detailed 
multi-epoch studies of the source at X-ray and ultraviolet (UV) wavelengths, and
found significant variations for the X-ray flux and broad UV absorption lines.

This source is relatively bright at the optical and MIR wavelengths, with
ZTF and CRTS magnitudes around 15 and WISE W1 and W2 magnitudes between 10--12.
Same as the above for J0122, the light curve data were obtained from these 
surveys' database.

There were seven X-ray observations of the source field, two each  
with {\it Chandra}, {\it XMM-Newton}, and 
{\it Neil Gehrels Swift Observatory (Swift)}, and one with 
{\it Nuclear Spectroscopic Telescope Array (NuSTAR)}. \citet{sco+15} analyzed
the {\it Chandra} and {\it XMM-Newton} data in detail, 
and \citet{luo+13} reported the analysis of the {\it NuSTAR} data.
We used their results and derived the unabsorbed fluxes in 0.5--10\,keV 
for the source, in which the tool {\tt PIMMS} was used. 
We analyzed the {\it Swift} data and obtained fluxes in the same energy
range; the details of the analysis are provided in Section~\ref{seca:xray}.
The X-ray flux results are summarized in Table~\ref{tab:xray}.

\subsection{J2131$-$1127}
This source was detected by CRTS, ZTF, and WISE. The CRTS light curve
is between 16--17\,mag, and the ZTF $zr$ is within the same
range but the $zg$ is notably 1-mag larger. The WISE W1 and W2 magnitudes
are around 13 and 12 respectively. The information for the data is
given in Table~\ref{tab:data}. 

There is no identification information for this source, but it was detected
by the Two Micron All Sky Survey in the infrared $JHK_s$ 
(2MASS; \citealt{skr+06}) 
and by WISE in the W1--W4 bands.
Both its MIR colors 
(W1$-$W2$\simeq$1.1, W2$-$W3$\simeq$2.9;
see, e.g., \citealt{wri+10}) and infrared colors (such as 
$J-K_s\simeq 0.77$, $K_s-$W3$\simeq5.5$; see, e.g., \citealt{tw13}) 
put the source in the quasar category.

There were six usable {\it Swift} XRT observations conducted in 2019 
covering the field of J2131$-$1127 (hereafter J2131). The exposure times 
were short, from 178\,sec to 1373\,sec, 
and the source was not detected in any of them. We used the online 
tools\footnote{https://www.swift.ac.uk/user\_objects/} for the data analysis. 
Using the data with the longest exposure (1373\,sec),
we derived an unabsorbed flux upper limit of 
6.2$\times 10^{-13}$\,erg\,s$^{-1}$\,cm$^{-2}$ in 0.3--10\,keV,
for which a power law was assumed with the hydrogen
column density and photon index fixed at the Galactic value 
4.6$\times 10^{20}$\,cm$^2$ \citep{hi4pi16} and 1.0, respectively.

There is also a $\gamma$-ray source, 4FGL~J2131.4$-$1124 in the
{\it Fermi} LAT catalog, possibly associated with J2131. 
The latter is $\simeq 5\arcmin$ 
away from the former 
and within the 95\%-confidence error 
circle of the former ($\sim 9\arcmin$). The $\gamma$-ray source
(unidentified or unassociated in the {\it Fermi} LAT catalog; \citealt{dr3})
is faint, having a total TS value of 46.4, and has emission described
with a power law ($\Gamma=2.81\pm0.15$). Similar to that for J0122, we 
constructed a 90-day
binned light curve of the source, but no obvious variability was seen 
in the light curve.

Based on the source's infrared colors and its possible association with
a $\gamma$-ray source, we suggest it as a candidate AGN. 
Observations for verification are needed. We note that 
given its optical brightness, if it is an AGN,
we would detect its X-ray emission around
a flux of $10^{-13}$\,erg\,s$^{-1}$\,cm$^{-2}$ 
(e.g., \citealt{he+19}).

\section{Periodicity analysis and results}
\label{sec:res}

The optical light curves of J0122, J1007, and J2131, as well as their
WISE W1 and W2 band magnitudes, are shown in Figures~\ref{fig:j01}, 
\ref{fig:j10}, and \ref{fig:j21}, respectively. As can be seen, the sources
all appear to have clearly visible modulation at the optical wavelengths, 
and even their WISE light curves show possibly related variations. We 
studied the modulation in each optical light curve and 
investigated the variations seen in the MIR bands. Below, we first
describe our periodicity analysis method 
(Section~\ref{subsec:met}), and then provide the results
from our studies of the optical light curves (Section~\ref{subsec:opt})
and MIR light curves (Section~\ref{subsec:mir}).

\subsection{Periodicity analysis method}
\label{subsec:met}

As the observed light curves of the three sources all show
sinusoidal-like modulation (Figures~\ref{fig:j01}, \ref{fig:j10}, 
\ref{fig:j21}) and the previous cases, i.e., those
reported such as in \citet{gra+15a} and \citet{liu+15}, were well described by a
simple sinusoidal function, we adopted the same model form to study the
periodicities. A sinusoidal function, $A\sin2\pi (t-B)/P + C$, was used,
where $t$ is time, $P$ is the period, and $A$, $B$, and $C$ are the amplitude,
the starting time (or the phase), and the constant magnitude, respectively.
However, because the CRTS data are in $V$ band and ZTF data are in $zg$ and 
$zr$ bands, possibly having different $A$ and $C$ values,
we first normalized each set of the light-curve data in 
a band with the sinusoidal function (to obtain $A$ and $C$). 
The $\chi^2$ fitting was used to find the best fits.
The obtained results are given in Table~\ref{tab:if}, in which
the reduced $\chi^2$ values
(=$\chi^2$/DoF, where DoF is the degrees of freedom)
are included to indicate the goodness of the fits.
As a check by comparison, we also provide the 
$\chi^2/{\rm DoF}$ values when fitting each set of data with a constant
(given in the lines of only parameter $C$ in Table~\ref{tab:if}). The comparison
shows significant improvements. For example, we may use
the Akaike Information Criterion (AIC) to check the improvements quantitatively,
$AIC=N\ln({\rm RSS}/N)+(2k+1)$, where RSS is the residual sum of squares, 
$N$ is the number of the data points, and $k$ is the parameter 
number \citep{bj17}. The $\Delta AIC$ value is $\simeq -126$ when we compare
the fitting results from the sinusoid model to those of the constant model
for the CRTS data of
J0122, and similar results can be obtained for the other data sets
given the large numbers of the data points. The large $\Delta AIC$
values indicate high signficances for the presence of the sinusoidal variations in 
the light curves (e.g., \citealt{rgb22} and references therein).

For the sinusoidal fitting, the $\chi^2/{\rm DoF}$ values are 
mostly in a
range of $\simeq$0.6--2.9. Here it should be noted that AGN 
generally show stochastic flux variations at a level of $<10\%$ 
(e.g., \citealt{van+04}), which implies that a systematic uncertainty may be
added to account for the variations in our fitting and
thus the $\chi^2/{\rm DoF}$ values could be further reduced.
However for J2131, the $\chi^2/{\rm DoF}$
values from fitting the ZTF $zg$ and $zr$ data sets are large,
$\simeq 8.8$ and 25 respectively (Table~\ref{tab:if}), the reason for
which is discussed below in Section~\ref{subsec:opt}.

With the obtained $A$ and $C$, we normalized the light-curve sets of a source
by subtracting $C$ from it and dividing it by $A$.  In this way, a 
normalized light curve of a source was constructed, with an average of 
$\sim$0 and a range of 1 to $-$1 (see the left bottom 
panels of Figures~\ref{fig:j01}, \ref{fig:j10}, and \ref{fig:j21}).
These light-curve data were then fitted with a sinusoidal function again, 
to mainly determine $P$ and $B$. Here we set $B$ as one parameter, since
the three optical bands have close wavelength values and we would not expect
phase shifts in these optical bands based on the current studies of SMBHB 
systems. 
In this last step, the Markov Chain 
Monte Carlo (MCMC) code {\tt emcee} \citep{fd+13} was used. 
\begin{table}
        \centering
       \caption{Initial fitting results to individual CRTS and ZTF light curves}
\begin{threeparttable}
        \begin{tabular}{lccccc}
                \hline
		Source    & Data     & $\chi^2/{\rm DoF}$ & DoF  & $A$       & $C^{\dagger}$     \\
                \hline
		J0122+1032  & CRTS-$V$  &2.39 & 129      &0.321   &19.314   \\
			   &    & 6.46 & 132 &    & 19.262	\\ 
		& ZTF-$zg$ &1.86  & 264     &0.489   &19.540       \\
			&   & 8.06 & 267 &  & 19.306	\\
		& ZTF-$zr$ &2.22  & 408     &0.394   & 19.051       \\
			&  & 13.22 & 411 &  &  18.876	\\
                \hline
		J1007+1248 & CRTS-$V$  & 0.59 & 174      & 0.168    &15.061  \\
			   &          &  3.57 & 177  &     & 15.145	\\
		 & ZTF-$zg$ & 2.75 & 178     &0.210   &15.159     \\
			&	&    147.21  & 181     &   & 15.101	\\
		& ZTF-$zr$ & 2.92  & 258     &0.192   &15.042     \\
			&	&   128.33   & 261     &  & 15.007	\\
\hline
		J2131-1127 & CRTS-$V$  &2.87 & 383    &0.151   &16.898   \\
			&	& 4.68   & 386 &	& 16.880	\\
		& ZTF-$zg$ & 8.77 & 148    & 0.366  &17.541            \\
			&	& 117.93 & 151 & 	& 17.485	\\
		& ZTF-$zr$ & 25.02 &168   & 0.381   &16.622          \\
			&  	& 500.97 & 171 &	& 16.616 \\
            \hline
        \end{tabular}
\begin{tablenotes}
%%\small 
\item $^{\dagger}$ Lines of only $C$ parameter are the fits to a constant.
\end{tablenotes}
\end{threeparttable}
        \label{tab:if}
\end{table}

\subsection{Optical light curve fitting results}
\label{subsec:opt}

From fitting the long-term light curves, which have a total time length of
approximately 6,000 days,
we obtained the periods of 2169.7, 2103.1, and 1462.6\,day for J0122, J1007,
and J2131, respectively. The other results are given in Table~\ref{tab:fits}.
The modulation of each source is reasonably well described by
each sinusoidal fit (shown in the left bottom panels
of Figures~\ref{fig:j01}, \ref{fig:j10}, and \ref{fig:j21}),
as the CRTS and ZTF light-curve parts for each source are smoothly 
connected by the fit
(note the two sets of data have a gap of $\sim$1600\,day).
The values of $A$ and $C$ are close to be 1 and 0 respectively, indicating
there were no significant mis-fit problems in the light-curve normalization
step; there could be such problems since the ZTF data do not cover 
a full cycle of a periodicity.

\begin{figure}
	\includegraphics[width=0.48\textwidth]{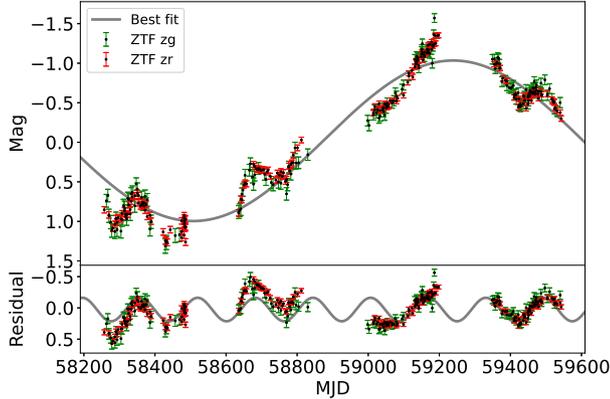}
	\caption{Sinusoidal fit (with $P=1462.6$ day) to the 
	ZTF $zg$ (green) and $zr$ (red) light curve parts of J2131
	({\it top} panel) and the residuals ({\it bottom} panel).
	Modulation with a period of $\simeq$162.2\,day and an 
	amplitude of $\simeq$0.188\,mag possibly exists in the light curve part,
	which is indicated by the black curve in the bottom panel.}
	\label{fig:res}
\end{figure}

We note that the $\chi^2/{\rm DoF}$ value for J2131 is again 
large (=10.5). Carefully examining the ZTF $zg$ and $zr$ light curve parts, 
wiggling around the model fit can be seen (Figure~\ref{fig:res}).
Interestingly, the residuals
of the light curve parts subtracted from the best fit suggest the existence
of short-term periodic modulation. We tested to fit the residuals 
with a sinusoidal function,
and obtained $P=162.17\pm0.18$\,day and $A=0.188\pm0.004$\,mag with
$\chi^2/{\rm DoF} = 13.2$. This large value reflects the significant
deviations of some data points from the short-term model fit 
(particularly visible at 
$\sim$MJD~59000). In any case, the reduced $\chi^2$ from the long-term
($P=1462.6$\,day)
fit for this part of the data is 21.2, indicating significant improvement
for describing the data variations. As the total $zg$ and $zr$ data points 
are 324,
$\Delta AIC\simeq -126$ is obtained when comparing the model fit having
the short periodicity included with that without it.
The residuals and the fitting results strongly suggest that there is another 
short-term periodicity, in addition to the 1462.6-day long periodicity.

\begin{table*}
        \centering
        \caption{Results from fitting the normalized light curves 
	of the three AGN.}
        \begin{tabular}{cccccccc}
                \hline
		Source     &$ \chi^2/{\rm DoF}$ & $A$                        &$B$                            &$C$                        & $P$ (day)                 \\
                \hline
                J0122+1032 &2.8    &$1.12_{-0.02}^{+0.02}$      &$693.2_{-209.8}^{+211.3}$   &$0.19_{-0.01}^{+0.01}$     &$2169.7_{-7.9}^{+7.7}$   \\
                J1007+1248 &2.9    &$1.036_{-0.005}^{+0.005}$      &$954.3_{-159.8}^{+157.0}$   &$0.060_{-0.005}^{+0.005}$     &$2103.1_{-5.7}^{+5.8}$\\
                J2131$-$1127 &10.5   &$1.015_{-0.004}^{+0.004}$      &$1098.9_{-95.5}^{+93.3}$    &$-0.019_{-0.003}^{+0.003}$    &$1462.6_{-2.4}^{+2.4}$   \\
                \hline
        \end{tabular}
        \label{tab:fits}
\end{table*}

\subsection{Studies of MIR variations}
\label{subsec:mir}
The WISE light curves of the three sources show variations, 
particularly clear and visible for J1007 and J2131 
(Figures~\ref{fig:j01}, \ref{fig:j10}, and \ref{fig:j21}). 
We calculated the reduced $\chi^2$ values of the W1 and W2 light curves
of the sources by fitting to a constant, and provided them
in Table~\ref{tab:fmir}. The $\chi^2/{\rm DoF}$ values for J0122 are $\geq 2.0$, and
for J1007 and J2131, the values are $\geq 5$. 
In particular, the light curves of J2131 show
signs of possible periodic modulation. We thus investigated the variations
of the light curves by fitting them with the same sinusoidal function 
given above, and the {\tt emcee} was used. From the fitting, when
all parameters were set free, the periods obtained are larger than those
of the optical light curves (see Table~\ref{tab:fmir}). Especially for J1007, 
the values
from the W1 or W2 bands are nearly 10 times or 5 times that of the optical 
one, which is because the whole light curves appear upward curved. 
The results indicate that the data can not provide good
constraints on periodicities, and no values, similar to those
optical ones, can be found. 
However, the fitting results do suggest modulation in the light
curves. 
For example, when we compare the fitting results from the sinusoid model
to those of the constant model,
for J0112 that has the smallest $\chi^2/{\rm DoF}$ values, $\Delta AIC$
are $\simeq -59$ and $\simeq -74$ in the W1 and W2 band respectively.

We then tested to fix the periods at the values of the optical 
ones in our fitting.
The $\chi^2/{\rm DoF}$ values from the fitting became slightly larger
(but still smaller than those from a constant model),
and the amplitudes were changed to be smaller (see Table~\ref{tab:fmir}).
Comparing the results with those from before, when all parameters for 
the sinusoidal function were set free, $\Delta AIC\simeq 20$ and 12,
for example, for J0122 in the W1 and W2 band respectively. 
This indicates the 
latter model is preferred. Similar conclusions can be made for J1007
and J2131.
Given the results of the comparison, no  
periodicities similar to the optical ones have been found.

\begin{table*}
        \centering
        \caption{Results from sinusoidal fitting to the WISE light curves of the three AGN}
	\begin{threeparttable}
        \begin{tabular}{ccccccc}
                \hline
		Source (Data)  & $\chi^2/{\rm DoF}$ & DoF &  $P$ (day)  & $A$       &$B$          & $C^{\dagger}$         \\
                \hline
		J0122+1032 (W1) & 2.4 &  183 & 2773.9 & 0.123      &1148.3     &14.752 \\
		(W1) & 2.7  & 184  & 2169.7$^\ast$  & 0.100     & 2730.8      &14.751        \\
		(W1) & 3.4 & 186 &		&	&	& 14.782 \\
		 (W2) & 1.3 & 183 & 3200.5 & 0.179      & 2228.8      &14.259 \\
		     (W2) & 1.4 & 184 & 2169.7$^\ast$ & 0.151   & 2954.1  &14.279\\
		(W2) & 2.0 & 186 & 		&	&	& 14.380 \\
		    \hline
		J1007+1248 (W1) & 1.6 & 192 & 19656.3 & 1.631      &72669.4     & 9.993 \\
		 (W1) &5.8 & 193 & 2103.1$^\ast$  &0.067  &3022.3 &11.575 \\
		(W1) & 11 & 195 & 		&	&	& 11.562 \\
		  (W2) & 1.5 & 192 & 9256.5 & 0.273      & 111314.1    &10.251 \\
		 (W2) & 2.4 & 193 & 2103.1$^\ast$  &0.048   &3066.7     &10.491   \\
		(W2)  & 5.0 & 195 & 		&	&	& 10.483 \\
		\hline
		J2131$-$1127 (W1) & 10.4 & 187 & 2178.8 & 0.221      &5710.0      &13.292 \\
		 (W1) &16.7 & 188 & 1462.6$^\ast$  &0.152  &2425.1 &13.256      \\
		(W1) & 26 & 190 &		&	&	& 13.283 \\
		(W2) & 7.1 & 187 & 2404.1 & 0.146   & 9953.7      &12.231 \\
		(W2) & 9.0 & 188 & 1462.6$^\ast$   & 0.102    &3866.0 &12.216     \\
		(W2) & 13 & 190 &		&	&	& 12.247\\
		\hline
        \end{tabular}
		\begin{tablenotes}
%	\centering{
\item $^\ast$ Periods are fixed at the best-fit values of the optical light curves.
\item $^\dagger$ Lines of only $C$ parameter are the fits to a constant. 
		\end{tablenotes}
	\end{threeparttable}
        \label{tab:fmir}
\end{table*}

\section{Discussion and Summary}
\label{sec:dis}

By exploring the ZTF data, we have found three good cases of long-term periodic
modulation. These three cases arise from a blazar (J0122), 
a Seyfert 1 galaxy (J1007), and a candidate AGN (J2131). Combining with 
the CRTS data, the periodicities were determined
by fitting the CRTS plus ZTF light curves of the three sources 
with a sinusoidal function. The periods determined for J0122, J1007, and J2131
are $\simeq$2170, 2103, and 1463\,day, respectively. While the periods are 
long, the total
lengths of the data for the sources are $\sim$6000\,day, covering relatively 
well the periodic modulation of each source. Based on the light-curve 
fitting 
results, we consider that the periodicities have been clearly revealed.

Such sinusoidal modulation seen in AGN has been widely discussed as evidence
for the existence of SMBHBs at the center of AGN. The modulation would reflect
the accretion rate changes according to numerical simulations for SMBHBs,
although how the observed optical modulation
is produced is not totally clear and under discussion (e.g., \citealt{gra+15a}).
One model that may work considers the relativistic Doppler effect of
emission from a close SMBHB \citep{dhs15}, which has been applied to 
explain AGN periodicities, quasi-periodicities, and aperiodic variations
(e.g., \citealt{yan+18,cha+18}).
Taking our cases as the example, the orbital
periods $P'_{\rm yr}$ (in units of year) at the host galaxies would 
be $\sim 6/(1+z)$ or $\sim 4/(1+z)$, and the separation
distances between two black holes would be 
$\simeq$0.002\,pc $(1+q)^{1/3} (M_1/10^8\,M_{\odot})^{1/3}(P'_{\rm yr})^{2/3}$, 
where $M_1$ is the mass of the primary black hole and
$q=M_2/M_1$ the mass ratio of the secondary (mass $M_2$) to the primary.
Assuming a circular orbit,
the velocity $v_2$ of the secondary is 
$v_2/c\simeq 0.1[1/(1+q)](M/10^9\,M_{\odot})^{1/3}(P'_{\rm yr})^{-1/3}$,
where $c$ is the speed of light and $M=M_1+M_2$ is the total mass. 
The observed flux $F$ would mainly arise from this secondary black hole
as it captures most of the matter coming from the circumbinary
disk, and given its relativistic velocity, $F$ would be orbitally 
modulated due to the Doppler effect
with an amplitude of $\delta F/F=(3-\alpha)(v_2/c)\sin i$,
where $\alpha$ is the 
power-law index for an emitted spectrum, and $i$ is the inclination angle
of the binary orbit (see details in \citealt{dhs15}).  

Since only for J1007, there is information of its redshift and estimated
black hole mass, we checked if the model could explain its modulation.
We de-reddened its SDSS spectrum with the Galactic extinction 
$E(B-V)= 0.033$ \citep{sfd98} and shifted the spectrum back 
to the rest wavelength with $z=0.24$. Fitting the $V$, $zg$, and $zr$ 
(for the latter two, see \citealt{bel+19}) wavelength ranges of 
the corrected spectrum, respectively, with a power-law 
$f_\nu\sim \nu^{\alpha}$, the $\alpha$ values were estimated to
be $\simeq 0.88$, 0.04, and $-0.16$. We then assumed $M=M_t$ and inserted
$P\simeq 5.76$\,yr (or $P'=4.64$\,yr), and obtained 
$\delta F/F \simeq \{0.16, 0.22, 0.23\}\times [1/(1+q)] (M/1.87\times 10^9)^{1/3} \sin i$
in the three bands. The amplitudes approximately match the observed values
(Table~\ref{tab:if}), although the $q$ and $i$ terms 
can significantly decrease the values if $q\sim 1$ or $i$ is small. 
We note that the model is rather simplified, such that
the optical continuum could consist of emission not only from
the secondary black hole (e.g., \citealt{dhs15,far+15}) and 
there are possible large 
uncertainties in the estimation---for example, the $\alpha$ values were 
estimated from a single spectrum which was taken before the times of the
CRTS data and long before those of the ZTF data. 
For J0122 and J2131, we can derive to find
$M=10^9\,M_{\odot} (\delta F/0.1F)^3 [(1+q)/(3-\alpha)]^3 (\sin i)^{-3} P'_{\rm yr}$. 
By inserting their $\delta F/F$ at $V$ band and $P=P'(1+z)$ values into 
the formula, we obtained
$M\sim \{2\times 10^{11}\,M_{\odot}$, $1.4\times 10^{10}\,M_{\odot}\} \times [(1+q)/(3-\alpha)]^3 (\sin i)^{-3}/(1+z)$ 
for the first and the latter respectively. 
In order not to have too large a mass (e.g., $\geq 10^{11}\,M_{\odot}$)
for the SMBHB in J0122, $q$ should be close to zero, $\alpha \sim 0$,
and $i$ should not be small.

%While no clear periodicities could be determined in the MIR light curves, according to the model proposed by \citet{dh17}, the dust torus surrounding an SMBHB could emit modulated infrared fluxes as the result of the reverberation of the modulated optical and UV emission. Among the three cases, J2131 showed the largest MIR variations and possible modulation pattern similar to its optical one (cf., Figure~\ref{fig:j21}). Taking the face values of the amplitudes from the sinusoidal fitting to the light curves with fixed $P=1462.6$\,day, the amplitude ratio of the MIR W1 (W2) to the optical modulation is $\sim$40\% (27\%). Note that the ZTF data are simultaneous to the WISE data in part, while the CRTS data are not (and the $V$ band amplitude is nearly equal to that of the W1 band). Also from the fitting, we found that the optical periodicity could lead the MIR ones in time by $t_d \sim (1300 +NP)$\,day (i.e., from the optical peak to the MIR peaks), where for simplicity no redshift is considered and $N$ is the number of additional cycles of the periodicity.  Based on the simplest, isotropic case considered in \citet{dh17}, the dust would have radius $R_d = c t_d \sim (1 + 1.2N)$\,pc. This radius value is in the size range estimated for AGN tori (e.g., \citealt{rr17}). However in order to produce the relatively large amplitude ratios,  the dust torus in J2131 would be required to be close to the central SMBHB, i.e., $N=0$ (see details in \citealt{dh17}).

For J2131, in addition to the $\sim$4-yr long periodicity,
modulation with a short period of $\simeq 162$\,day possibly also existed.
This phenomenon, 
if verified with
continuous observations, would be an intriguing feature to investigate.
We suspect that it could be explained with a hot spot at the outer edge
of the mini-disc around the secondary black hole. The gravitational radius
is $r_g = GM_2/c^2 \simeq 1.5\times 10^{13}$\,cm for a 10$^8\,M_{\odot}$ 
black hole.
At radius $r\simeq 270 r_g$ (or $\simeq 0.001$\,pc), 
the Keplerian orbital period is $\simeq 162$\,day. As the AGN disc sizes are
found to be mostly in a range of $10^{-3}$--$10^{-2}$\,pc (possibly depending
on the black hole mass; e.g., \citealt{jha+22}), a hot spot that orbits
the SMBH and is located at the outer edge of the SMBH's disc could
give rise to the short period signal.
Such a hot spot could be produced by the interaction between
the stream of matter from the circumbinary disk with the outer edge
of the mini-disc (e.g., \citealt{hms07,rkm14}). We realize that there could 
be alternative
explanations for the presence of two periodicities in J2131, such as the short 
periodicity being a jet's wobbling timescale \citep{lis+18}. 
As such discussion would deviate away from
the focus of this work, we refer to \citet{zha+22} for the discussion 
of alternative scenarios.
\begin{figure}
	\includegraphics[width=0.48\textwidth]{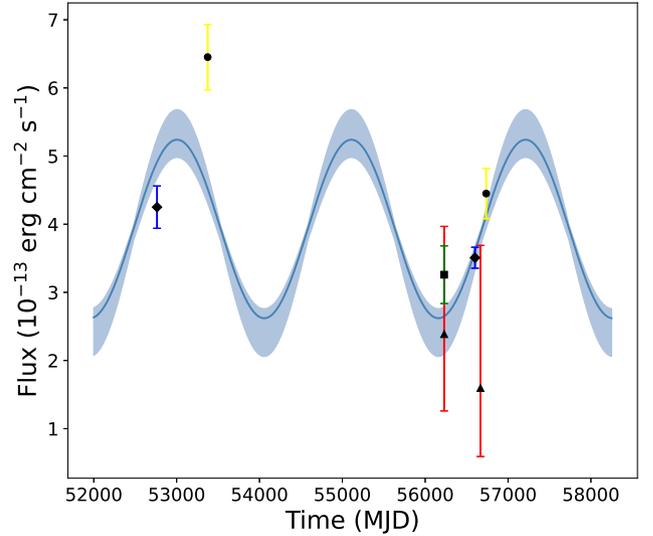}
	\caption{Unabsorbed 0.5--10 keV fluxes of J1007 
	(diamonds: {\it XMM-Newton}; circles: {\it Chandra}; square: {\it NuSTAR}; triangles: {\it Swift}).
	The flux variations
	are possibly consistent with the optical periodicity (blue line).}
	\label{fig:xray}
\end{figure}

Finally, \citet{sco+15} noted that J1007 showed upto 40\% X-ray flux 
variations, and the variations were likely intrinsic because
no significant variations in the amount of X-ray absorption were seen.
We plotted all the unabsorbed 0.5--10 keV fluxes from Table~\ref{tab:xray}
in Figure~\ref{fig:xray}, 
and tested to fit the fluxes with a sinusoidal function, but with
the parameters $P$ and $B$ fixed at the values of the determined optical 
periodicity. As shown in Figure~\ref{fig:xray}, the X-ray flux variations
are possibly consistent with the optical periodicity, which may suggest 
that
the X-ray flux variations are actually caused by the presence of an SMBHB
\citep{ser+20}; 
the accretion from the circumbinary disc to the black holes is
orbitally modulated. 
This possibility predicts
very certain X-ray flux variations and can be verified with further
{\it Chandra} or {\it XMM-Newton} observations.

As a summary, we have discovered three cases of optical modulation in AGN,
and the periodicities are likely caused by the widely-discussed SMBHB
scenario. For J1007 that has redshift and estimated black hole mass,
we have found that the modulation could be explained with the Doppler effect
of emission from a secondary SMBH with a relativistic orbital velocity.
We have also analyzed the MIR light curve variations for the three AGN
and found that J2131 was the more variable one, but whether 
the variations are related to the optical ones is not clear.
%Its MIR variations were possibly periodic with the optical period, which would be theoretically expected as due to the dust reverberation of the modulated optical emission.
For J2131, its optical light curve showed additional variations around
its long-term modulation, for which a short periodicity was estimated,
and may be explained as the presence of a hot spot at the outer edge of the
mini-disc of the secondary black hole.
Also for J1007, which has been observed multiple times at X-rays, 
its X-ray flux varied possibly
with its optical periodicity. The verification of this possibility
would strengthen its case as a candidate SMBHB and suggest an accompanying 
window at X-rays for finding candidate SMBHBs \citep{ser+20}.
As several large optical transient survey
programs are or will be in operation, more data will certainly be collected 
for the three sources. We will be able to
keep checking the persistence of the periodicities, 
which could potentially establish them as the good SMBHB candidates.

\section*{Acknowledgements}

This work was
based on observations obtained with the Samuel Oschin Telescope 48-inch and 
the 60-inch Telescope at the Palomar Observatory as part of the Zwicky 
Transient Facility project. ZTF is supported by the National Science 
Foundation under Grant No. AST-2034437 and a collaboration including Caltech, 
IPAC, the Weizmann Institute for Science, the Oskar Klein Center at
Stockholm University, the University of Maryland, Deutsches 
Elektronen-Synchrotron and Humboldt University, the TANGO Consortium of 
Taiwan, the University of Wisconsin at Milwaukee, Trinity College Dublin, 
Lawrence Livermore National Laboratories, and IN2P3, France. Operations are 
conducted by COO, IPAC, and UW.

This work made use of data products from the Wide-field Infrared Survey 
Explorer, which is a joint project of the University of California, Los 
Angeles, and the Jet Propulsion Laboratory/California Insitute of Technology, 
funded by the National Aeronautics and Space Administration.

We thank the anonymous referee for very helpful comments.
Y. Luo has helped our understanding of SMBHB evolution scenarios, and
M. Gu about the origin of the AGN optical variability.
This research is supported by the Basic Research Program of Yunnan Province 
No. 202201AS070005, the National Natural Science Foundation of China 
No. 12273033, and the Original Innovation Program of the Chinese Academy of 
Sciences (E085021002).

%%%%%%%%%%%%%%%%%%%%%%%%%%%%%%%%%%%%%%%%%%%%%%%%%%
\section*{Data Availability}
The data underlying this article will be shared on reasonable request to
the corresponding author.



\bibliographystyle{mnras}

\appendix 
\section{X-ray observations of J1007+1248} 
\label{seca:xray} 

The two {\it Swift} observations of the source field was conducted with XRT 
in photon counting mode. We used the online 
tools\footnote{https://www.swift.ac.uk/user\_objects/} for the data analysis 
and source detection \citep{epo+20}.  In the first observation, the source was 
only found to have a $\geq 3\sigma$ detection significance by comparing 
its count rate (4.7$^{+3.1}_{-2.2}\times 10^{-3}$\,cts\,s$^{-1}$) with 
that of the expected background \citep{ebp+07,ebp+09}. Its unaborbed flux 
in 0.5--10 keV was estimated using the {\tt PIMMS} tool, where the Galactic 
$N_{\rm H}$ value 3.5$\times 10^{20}$\,cm$^{-2}$ \citep{hi4pi16} and 
photon index 1.0 were assumed. 
In the second observation, 
only 6 photons from the source were detected. A spectrum was created 
following \citet{ebp+09} and fitted with a power-law model (in which
the same Galactic $N_{\rm H}$ value was used). The corresponding 0.5--10\,keV
unabsorbed flux was obtained.
The fluxes from these two observations are given in Table~\ref{tab:xray}.

\begin{table*} 
	\centering 
	\caption{X-ray observations of J1007+1248 and flux measurement results} \begin{tabular}{lccccc} 
	\hline 
	Observation & Date  & Obsid & Exposure &  $F^{\rm unabs}_{0.5-10}/10^{-13}$  \\ 
	&  &      & (ks)          &           (erg\,cm$^{-2}$\,s$^{-1}$) \\ 
		\hline {\it Chandra}1    & 2005-01-05 & 5606       & 41 &  6.45$\pm$0.48          \\ 
		{\it Chandra}2    & 2014-03-20 & 16034      & 61 &  4.45$\pm$0.37            \\ 
		{\it XMM-Newton}1 & 2003-05-04 & 0140550601 & 22 &  4.25$\pm$0.31  \\ 
		{\it XMM-Newton}2 & 2013-11-05 & 0728980201 & 66 &  3.51$\pm$0.15          \\ 
		{\it Swift}1      & 2012-10-29 & 00080031001& 2.0       & $2.4^{+1.6}_{-1.1}$            \\ 
		{\it Swift}2      & 2014-01-10 & 00033077001& 1.9  & 1.6$^{+2.1}_{-1.0}$    \\ 
		{\it NuSTAR} & 2012-10-29 & 60001112002 & 33 &  $3.26\pm0.42$ \\
	\hline 
	\end{tabular} 
\label{tab:xray} 
\end{table*}

%%%%%%%%%%%%%%%%%%%%%%%%%%%%%%%%%%%%%%%%%%%%%%%%%%

% Don't change these lines
\bsp	% typesetting comment
\label{lastpage}
\end{document}